%% file: main.tex
\pdfoutput=1 

\documentclass[5p,times]{elsarticle}

\usepackage{algorithmic}
\usepackage{graphicx}
\usepackage{textcomp}
\usepackage{xcolor}
\usepackage{hyperref}
\usepackage{booktabs}
\usepackage[acronym,toc,shortcuts]{glossaries}
\usepackage{tikz}
\usetikzlibrary{quantikz}
\usepackage{adjustbox}
\usepackage{subcaption}
\usepackage{svg}
\usepackage{soul}
\usepackage{multirow}
\usepackage{listings}
\usepackage{comment}
\usepackage{pifont}% http://ctan.org/pkg/pifont
\usepackage{pgfplots}
\usepackage{enumitem}
\pgfplotsset{compat=1.18}

\newcommand{\cmark}{\ding{51}}%
\newcommand{\xmark}{\ding{55}}%

\definecolor{lightgray}{gray}{0.9}
\definecolor{shadecolor}{gray}{0.92}
\definecolor{dkgreen}{rgb}{0,0.6,0}
\definecolor{gray}{rgb}{0.5,0.5,0.5}
\definecolor{mauve}{rgb}{0.58,0,0.82}

\input{acronyms}
  
\usepackage{titlesec}

\titleformat{\paragraph}
{\normalfont\normalsize\itshape}{\theparagraph}{0.2em}{\normalfont\normalsize\itshape}
\titlespacing{\paragraph}
{0pt}{*1}{*1}[0.2pt]
\begin{document}

\begin{frontmatter}

\title{NetQIR: An Extension of QIR for Distributed Quantum Computing}

\author[citius]{F. Javier Cardama}
\author[citius,cesga]{Jorge Vázquez-Pérez}
\author[citius,dec]{César Piñeiro}
\author[citius,dec]{Tomás F. Pena}
\author[citius,dec]{Juan C. Pichel}
\author[cesga]{Andrés Gómez}

\affiliation[citius]{organization={Centro Singular de Investigación en Tecnoloxías Intelixentes (CiTIUS)},%Department and Organization
            addressline={Universidade de Santiago de Compostela}, 
            city={Santiago de Compostela},
            postcode={15782}, 
            %state={A Coruña},
            country={Spain}}
\affiliation[dec]{organization={Departamento de Electrónica e Computación},%Department and Organization
            addressline={Universidade de Santiago de Compostela}, 
            city={Santiago de Compostela},
            postcode={15782}, 
            %state={A Coruña},
            country={Spain}}
\affiliation[cesga]{organization={Galicia Supercomputing Center (CESGA)},%Department and Organization
            addressline={Avda.\ de Vigo S/N}, 
            city={Santiago de Compostela},
            postcode={15705}, 
            %state={A Coruña},
            country={Spain}}

%%%%%%%%%%%%%%%%%%%%%%%%%%%%%%%%%%%%%%%%

\begin{abstract}

The rapid advancement of quantum computing has highlighted the need for scalable and efficient software infrastructures to fully exploit its potential. Current quantum processors face significant scalability constraints due to the limited number of qubits per chip. In response, distributed quantum computing (DQC) ---achieved by networking multiple quantum processor units (QPUs)--- is emerging as a promising solution. To support this paradigm, robust intermediate representations (IRs) are needed to translate high-level quantum algorithms into executable instructions suitable for distributed systems. This paper presents NetQIR, an extension of Microsoft's Quantum Intermediate Representation (QIR), specifically designed to facilitate DQC by incorporating new instruction specifications. NetQIR was developed in response to the lack of abstraction at the network and hardware layers identified in the existing literature as a significant obstacle to effectively implementing distributed quantum algorithms. Based on this analysis, NetQIR introduces new essential abstraction features to support compilers in DQC contexts. It defines network communication instructions independent of specific hardware, abstracting the complexities of inter-QPU communication. Leveraging the QIR framework, NetQIR aims to bridge the gap between high-level quantum algorithm design and low-level hardware execution, thus promoting modular and scalable approaches to quantum software infrastructures for distributed applications. Furthermore, its design may serve as a foundational component for future implementations of distributed quantum standards such as the Quantum Message Passing Interface (QMPI).
\end{abstract}


\end{frontmatter}

\input{0_Introduction}
\input{1_Related_Work}
\input{2_Layered_Abstraction_Model}

\input{3_NetQIR}

\input{4_Results}

\input{5_Conclusions}

%%%%%%% -- PAPER CONTENT ENDS -- %%%%%%%%

{\small
\section*{Declaration of generative AI and AI-assisted technologies in the writing process}

During the preparation of this work the authors used ChatGPT in order to improve language and readability. After using this tool, the authors reviewed and edited the content as needed and take full responsibility for the content of the publication.
}

{\small
\section*{Acknowledgments} 

This work was supported by MICINN through the European Union NextGenerationEU recovery plan
(PRTR-C17.I1), the Galician Regional Government through "Planes Complementarios de
I+D+I con las Comunidades Autónomas'' in Quantum Communication, MINECO (grants PID2019-104834GB-I00, PID2022-141623NB-I00 and PID2022-137061OB-
C22), Consellería de Cultura, Educación e Ordenación Universitaria Galician Research Center accreditation 2024-2027 ED431G-2023/04, and the European Regional Development Fund (ERDF). %, which acknowledges the CiTIUS-Research Center in Intelligent Technologies of the University of Santiago de Compostela as a Research Center of the Galician University System.
}

%%%%%%%%% -- BIB STYLE AND FILE -- %%%%%%%%
%\bibliographystyle{IEEEtranS}
\bibliographystyle{elsarticle-num}
\bibliography{externmendeley, refs}
%\input{main.bbl}
%%%%%%%%%%%%%%%%%%%%%%%%%%%%%%%%%%%%

\end{document}

%% file: acronyms.tex
\newacronym{fpga}{FPGA}{field-programmable gate array}
\newacronym{gpu}{GPU}{graphics processing unit}
\newacronym{hpc}{HPC}{high-performance computing}
\newacronym{dqc}{DQC}{distributed quantum computing}
\newacronym{qpu}{QPU}{quantum processing unit}
\newacronym{ir}{IR}{intermediate representation}
\newacronym{epr}{EPR}{Einstein-Podolsky-Rosen}
\newacronym{qnpu}{QNPU}{quantum network processing unit}

\newacronym{openmp}{OpenMP}{Open Multi-Processing}
\newacronym{mpi}{MPI}{Message Passing Interface}
\newacronym{qmpi}{QMPI}{Quantum Message Passing Interface}
\newacronym{qasm}{QASM}{Quantum Assembly Language}
\newacronym{nisq}{NISQ}{Noisy Intermediate-Scale Quantum}
\newacronym{dqcs}{DQCS}{Distributed Quantum Computing Simulator}
\newacronym{qlan}{QLAN}{Quantum Local Area Network}
\newacronym{qir}{QIR}{Quantum Intermediate Representation}
\newacronym{qft}{QFT}{Quantum Fourier Transformation}

\newacronym{sdk}{SDK}{Software Development Kit}

%% file: 0_Introduction.tex
\section{Introduction}
The evolution of computing has progressed from simple mechanical calculators to modern-day classical computers, that have significantly transformed numerous fields, including science, engineering, and everyday life. Despite these advances, classical computers face limitations in solving certain complex problems efficiently, such as factoring large numbers, simulating quantum systems, or optimising large-scale systems~\cite{Feynman1982, Markov2014}. This has led to the emergence of quantum computing, which leverages the principles of quantum mechanics to process information in fundamentally new ways, offering the potential to solve these intractable problems more efficiently than classical computers can achieve~\cite{buhrman1998quantum, Riel2021}.

Over the last few years, the development of a comprehensive software stack for quantum computing has gained importance in allowing the programming of quantum devices in a scalable and easy way. This software stack includes quantum high-level languages, compilers, and runtime environments designed to enable the programming and execution of quantum algorithms on quantum devices~\cite{Koen2020, Haner2021}. High-level quantum programming languages such as Q\#~\cite{Svore2018}, Quipper~\cite{Green2013}, or Qiskit~\cite{Aleksandrowicz2019} facilitate the development of quantum algorithms by abstracting the complexities of quantum hardware~\cite{Serrano2022}.

For the efficient execution of these algorithms, quantum code compilers play a crucial role. A compiler is a software program that translates high-level languages into low-level instructions that quantum processors can execute~\cite{aho-compilers-book}. In classical computing, the concept of \gls*{ir} was introduced as an abstract-machine code to facilitate the development of new compilers~\cite{Stanier2013}. This concept was extended in the world of quantum computing to allow a common \gls*{ir} as an intermediate step between high-level and back-end languages. The main objective of using an \gls*{ir} is to facilitate the optimisation of quantum codes and, simultaneously, to ensure their compatibility with different hardware backends~\cite{Metodi2010, Hietala2019}.

%Despite these advances, 
One of the critical challenges in quantum computing remains the scalability and noise of the qubits. Current quantum hardware is limited by the number of qubits that can be reliably maintained and manipulated on a single chip, thus complicating the development of more complex algorithms~\cite{Almudever2017}. These limitations have led to the development of new computing approaches, one of which is the design of modular architectures based on \gls*{dqc}. In these architectures, multiple \glspl*{qpu} are networked together to work on a problem collaboratively~\cite{Beals2013, Loke2022, Wakizaka2023}. \gls*{dqc} uses both quantum and classical communications to distribute and synchronise computations across \glspl*{qpu}, thereby potentially overcoming the scalability constraints of individual quantum chips~\cite{Meter2009, Rodrigo2020, Barral2024}.
%\gls*{dqc} in modular architectures where multiple \glspl*{qpu} are networked together to work on a problem collaboratively~\cite{Beals2013, Loke2022, Wakizaka2023}. \gls*{dqc} uses both quantum and classical communications to distribute and synchronise computations across \glspl*{qpu}, thereby potentially overcoming the scalability constraints of individual quantum chips~\cite{Meter2009, Rodrigo2020, Barral2024}.

\ac*{dqc} introduces a level of complexity over monolithic quantum computing systems (systems consisting of a single QPU), especially to the requirements of quantum networking. Because of this, an abstraction model of the complete process of a DQC algorithm, from its high-level specification to the specific back-end where it is supposed to run,  must be defined to facilitate the development of tools according to their specific use. Additionally, this need highlights the importance of defining an \ac*{ir} that not only enables the efficient programming of quantum algorithms within this abstraction framework but also addresses the specific challenges inherent to \ac*{dqc}, such as optimized communication protocols and precise synchronization mechanisms across QPUs~\cite{chong2017programming}. 

In the context of \ac*{dqc}, it is essential to distinguish it from the concept of the Quantum Internet~\cite{wehner-quantum-internet-survey-1, illiano-quantum-internet-survey-2}. While \ac*{dqc} relies on a quantum interconnection network, which falls within the scope of the Quantum Internet, its primary focus remains on computation rather than networking itself~\cite{azuma2023quantumrepeaters}. Therefore, it is crucial to abstract away the complexities of quantum networking and, instead, define the appropriate high-level instructions within an \gls*{ir} to ensure its efficient utilization. Currently, a wide range of tools are available for simulating quantum communication networks (network-level simulators)~\cite{bel2024simulatorsquantumnetworkmodelling}, such as SquidASM~\cite{Dahlberg2022} and Simulaqron~\cite{dahlberg2018simulaqron}, as well as discrete event simulation for quantum communication at the physical level, including NetSquid~\cite{coopmans2021netsquid} and SeQuence~\cite{wu2020sequence}. There are also simulators specifically designed for distributed quantum computing, such as QuNetSim, which follows a more point-to-point model, with network simulation handled by the in-process EQNS simulator~\cite{diadamo2021qunetsim}.

In response to the problems encountered in the literature, this paper proposes two main contributions to the state of the art:

\begin{itemize}
    \item The definition of a \textbf{layered abstraction model} for the correct implementation of IRs for DQC by collecting information from the literature, both from standards followed in classical computing and from attempts at quantum computing. This objective will allow other developers to implement new IRs following a model whose efficiency will be demonstrated later.
    \item The proposal of an \gls*{ir} that implements the abstraction layer model proposed in the previous objective. Our proposal, NetQIR, extends Microsoft’s QIR~\cite{QIRSpec2021}, augmenting it with advanced communication and distributed computation directives 
    to support interoperability and scalability, thereby facilitating robust quantum algorithm development in distributed quantum environments and hybrid programming.
\end{itemize}

\begin{figure*}[t!]
    \centering
    \begin{subfigure}[b]{0.49\textwidth}
        \centering
        \includegraphics[width=0.80\textwidth]{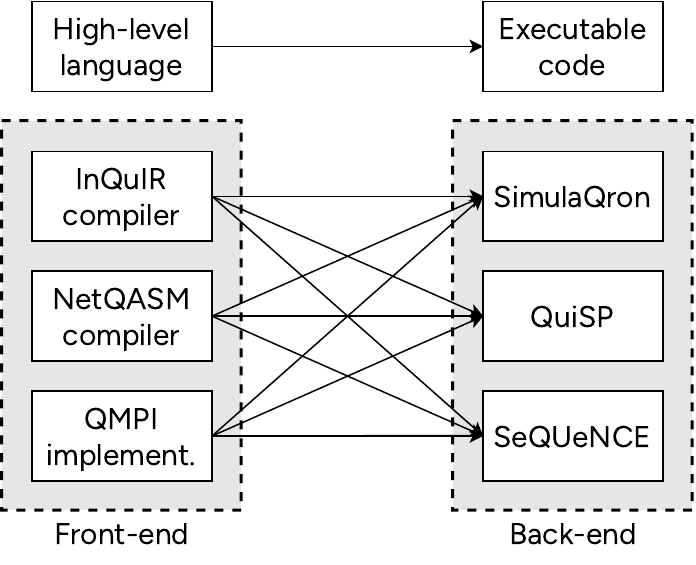}
        \caption{Without \gls*{ir} usage.}
        \label{subfig:intro:noIR}
    \end{subfigure}
    \hfill
    \begin{subfigure}[b]{0.49\textwidth}
        \centering
        \includegraphics[width=0.80\textwidth]{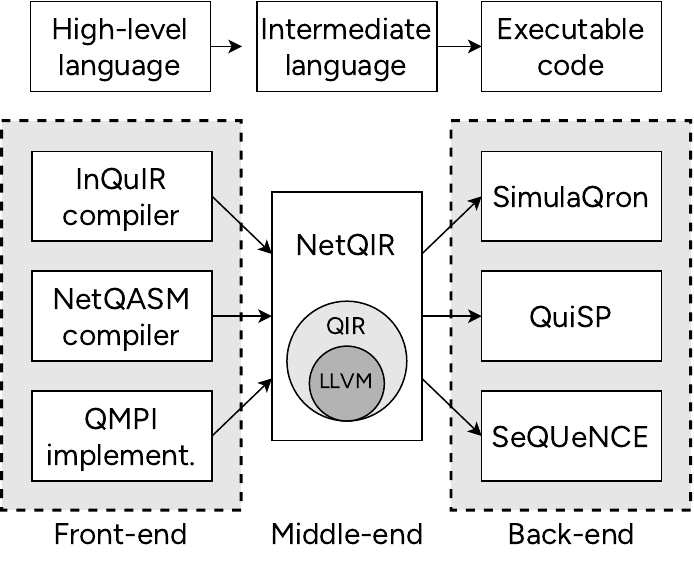}
        \caption{Applying NetQIR as \gls*{ir}.}
        \label{subfig:intro:IR}
    \end{subfigure}
    \caption{Comparison between the integration of NetQIR as \ac*{dqc} \ac*{ir} in a compilation scheme between \ac*{dqc} programming frameworks and quantum simulator backends.}
    \label{fig:intro:ir_example}
\end{figure*}

Figure~\ref{fig:intro:ir_example} illustrates the advantages of using NetQIR as an \gls*{ir} for DQC, highlighting its extension of QIR and, consequently, LLVM. This figure illustrates DQC tools compiled to various quantum network simulators. In the initial approach, without the use of IRs, the number of required compilers is $n \times m$. However, by leveraging an \gls*{ir} such as NetQIR, this complexity is significantly reduced to $n+m$, streamlining the compilation process\footnote{With $n$ as the number of \textit{front-ends} and $m$ of \textit{backends}.}.

The paper is structured as follows: initially, Section~\ref{sec:related-work} reviews the related work, focusing on existing \acp*{ir} and programming languages for \ac*{dqc}. Then, Section~\ref{sec:layered_abstraction_model} introduces a layered abstraction model, essential for \ac*{dqc}. It presents the development and network layers, detailing the specific components required to achieve a modular and interoperable architecture for DQC. In addition, Section~\ref{sec:netqir} presents the topics related to the proposed \gls*{ir}, including its specification and the tools developed to facilitate the design of new software for future developers. Subsequently, Section~\ref{sec:results} presents a comparative analysis of NetQIR in relation to other \acp*{ir} and \ac*{dqc} languages in a qualitative form; this qualitative specification will be done on the basis of the proposed abstraction layer model, citing the main reason why it is an important contribution to develop an \gls*{ir} like NetQIR. On the other hand, the need for a layered abstraction model is also quantitatively evaluated using real cases of \ac*{dqc} algorithms with different communication models and topologies, thus also demonstrating the quality of the qualitative evaluation. Finally, Section~\ref{sec:conclusions} concludes the work and specifies future work.

%% file: 1_Related_Work.tex
\section{Related Work}\label{sec:related-work}

In \gls*{dqc}, the software stack lacks available tools. From high-level languages to lower-level representations---and even development libraries---the literature offers few possibilities, as demonstrated by the state-of-the-art review conducted by Barral et al.~\cite{Barral2024}. This becomes even more evident when compared to the monolithic case, in which numerous programs, libraries and other software are available for developing quantum applications.

Focusing on the \glspl*{ir} in the monolithic quantum computing case, MLIR~\cite{Lattner2021MLIR,mccaskey2021mlir} or SQIR~\cite{hietala2021squir} are found, along with QIR~\cite{QIRSpec2021}, backed by the QIR alliance\footnote{\url{https://www.qir-alliance.org/}} ---from now on, it will be referred to as QIR---.
%We will highlight only one \gls*{ir} from monolithic quantum computing. It represents the core and the starting point of our work: Microsoft's QIR~\cite{QIRSpec2021}, backed by the QIR alliance\footnote{\href{https://www.qir-alliance.org/}{https://www.qir-alliance.org}} ---from now on, it will be referred to as QIR---. 
The latter is based on the LLVM \gls*{ir}~\cite{llvm} in an attempt to integrate quantum computation into the LLVM infrastructure\footnote{LLVM is a versatile framework for building compilers and code transformation tools. It lets developers write high-level language code that can be efficiently compiled into machine code for various architectures, with extensive code optimisation and analysis support.}. In fact,  QIR aims to integrate quantum directives with the classical compilation stack, leveraging the advanced LLVM tools to facilitate the generation of highly efficient quantum instructions. In this work QIR will be extended and, therefore, the LLVM \gls*{ir} will be further extended by introducing the necessary directives to perform quantum communications. Throughout this manuscript, this extension will be explained and exemplified.

For \gls*{dqc}, the two most popular specifications in the literature are InQuIR~\cite{Nishio2023} and NetQASM~\cite{Dahlberg2022}. The first one, InQuIR, is developed from the starting point solely as an \gls*{ir} for \gls*{dqc}. Their primary motivation stemmed from the absence of a dedicated \gls*{ir} for distributed quantum systems. InQuIR stands out for formally defining the grammar of the \gls*{ir}. Using this formalism, InQuIR defines the operational semantics of the \gls*{ir}, which allows it to define and predict how the InQuIR programs will behave under several circumstances. They even propose some important examples, such as deadlocks and qubit exhaustion, and a roadmap for solving these inconveniences. But InQuIR has several drawbacks: on the one hand, the syntax definition is quite inconsistent, mixing quantum and classical data without differentiating, and, on the other, it provides a too low-level approach with explicit generation of the \gls*{epr} pairs and instructions that acknowledge the architecture of the machine, which should be left to the backend, not to the \gls*{ir}. 

% InQuIR implementation also includes a toy compiler, which, when given a QASM code and an architecture, returns the corresponding InQuIR code. This exemplifies the abstraction problem that InQuIR presents: an \gls*{ir} should be platform-independent.

As an alternative, as mentioned, NetQASM~\cite{Dahlberg2022} presents an abstract architecture model composed of an application layer, which is responsible for the classical communications between nodes, and a so-called \gls*{qnpu}, which handles quantum computations and communications. This highlights the scope of NetQASM: the Quantum Internet. It is specifically designed for quantum networks, setting aside \emph{inter-core} communication, which does not require the additional layers that NetQASM introduces. Moreover, NetQASM presents a basic language, called \emph{vanilla}, and a set of variations specially designed for the different quantum architectures, called \emph{flavours}. The authors state that the vanilla version acts as an \gls*{ir}, and the different flavours act as assemblies. The main disadvantage of NetQASM, like \gls*{ir} for \gls*{dqc}, is that its architecture is too network-oriented rather than computation-oriented. It also does not consider conditional gates, which are constantly employed in quantum communication protocols, as part of the \gls*{ir}. What is actually done is to perform a measurement, send the result to the application layer and wait until the application layer returns a subroutine with the gate ---if the measurement was 1--- or without the gate ---in the opposite case---.

After discussing the state-of-the-art \glspl*{ir}, it is important to note that NetQIR will not extend QIR arbitrarily. Instead, it is designed to follow a process analogous to that of a well-known classical protocol, such as \gls*{mpi}. By using this protocol as a reference, the intention is to leverage its maturity to avoid repeating similar mistakes in the development process. A similar approach is adopted by \gls*{qmpi}, as proposed by Haner et al.~\cite{Haner2021}. As its name implies, it is an adaptation of the classical \gls*{mpi}~\cite{forum1994mpi} for quantum communications, achieved by defining analogous point-to-point and collective operations for the quantum pipeline. As this work will show, the difference between \gls*{qmpi} and NetQIR lies in two main pillars. First, \gls*{qmpi} functions as a high-level standard, much like \gls*{mpi}, whereas NetQIR serves as an \gls*{ir}. Second, while both \gls*{qmpi} and NetQIR define a set of directives, they differ slightly in their approach. \gls*{qmpi} establishes operations that closely mirror those of classical \gls*{mpi}, whereas NetQIR identifies certain issues with this strategy and opts to deviate somewhat from the classical model.

%defining point-to-point and collective quantum operations, and, concerning the latter, it introduces certain approaches that conflict with the non-cloning theorem. The fact that arbitrary quantum data cannot be copied eliminates the possibility of having collective operations such as a quantum broadcast, which cannot be implemented following this scheme because it would not be allowed to send a copy of the quantum data to each node~\cite{Van_Meter_2007}. In this work, we present the NetQIR collective operations and define an instruction similar to the broadcast introduced by \gls*{qmpi}, but without the need for copies.

In summary, both InQuIR and NetQASM exhibit certain aspects that may represent drawbacks for an \gls*{ir}. Additionally, \gls*{qmpi} defines a high-level standard that is strongly coupled with the classical \gls*{mpi}, introducing various complications and intricacies. NetQIR seeks to address these issues and this ma\-nus\-cript will detail the approach taken to achieve that goal and justify the decisions made in order to do so.

%\textcolor{red}{JCP: no se hace referencia a la Figura 1 en el texto}

% \begin{figure*}[t!]
%     \centering
%     \begin{subfigure}[b]{0.49\textwidth}
%         \centering
%         \includegraphics[width=0.8\textwidth]{/netqir-0.pdf}
%         \caption{Front-end to back-end scheme.}
%         \label{subfig:sec1:NetQIR-0}
%     \end{subfigure}
%     \hfill
%     \begin{subfigure}[b]{0.49\textwidth}
%         \centering
%         \includegraphics[width=0.8\textwidth]{/netqir-1.pdf}
%         \caption{NetQIR as middle-end in the scheme.}
%         \label{subfig:sec1:NetQIR-1}
%     \end{subfigure}
%     \caption{Comparison between the integration of NetQIR as \ac*{dqc} \ac*{ir} in a compilation scheme between \ac*{dqc} programming frameworks and quantum simulator backends.}
%     \label{fig:sec1:NetQIR}
% \end{figure*}

%% file: 2_Layered_Abstraction_Model.tex
\section{Layered Abstraction Model for Distributed Quantum Computing}\label{sec:layered_abstraction_model}

Software architectures nowadays strongly rely on abstraction mechanisms as a core principle. These simplify the development of new algorithms, platforms, compilers and tools. And \ac*{dqc} architectures are no different. This provokes that, similar to the monolithic---and even classical---case, creating a new \ac*{ir} for this paradigm should be hardware-independent and compatible with various quantum computing platforms. Therefore, it is necessary to define a layered abstraction model before developing the \ac*{dqc} \ac*{ir}.

\begin{figure}[t!]
    \centering
    \includegraphics[width=\linewidth]{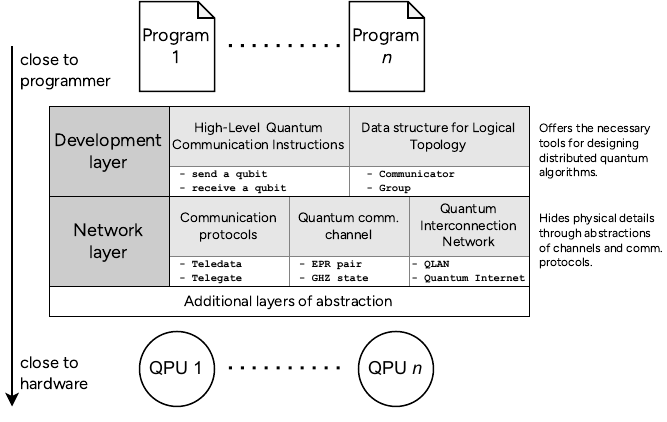}
    \caption{Abstraction layers relevant to the development of an abstract \ac*{ir} for \ac*{dqc}.}
    \label{fig:sec2:layered_abstraction_model}
\end{figure}

Figure~\ref{fig:sec2:layered_abstraction_model} depicts the abstraction layers relevant for executing algorithms in a \gls*{dqc} environment. This paper focuses on the computational part of distributed quantum, Therefore, two key layers are defined in our proposed model: the development layer and the network layer. The development layer provides users the necessary tools to design and implement \gls*{dqc} algorithms and software. Meanwhile, the network layer acts as an interface between the development layer and the quantum interconnection network, ensuring seamless interaction while managing its underlying characteristics. Additionally, the network layer interacts with lower layers as needed, further abstracting the physical complexities of the quantum network.

\subsection{Network layer}
As discussed above, the network layer aims to abstract the particularities of the quantum interconnection network to the development layer. For this purpose, it is necessary to identify this layer's main components, which are the \emph{quantum interconnection network}, the \emph{quantum communication channel} and the \emph{communication protocols}.
%In the next subsections, the details of each of these components are presented, explaining their importance and how they interact with each other.
%One of the most important is the \textbf{quantum interconnection network}, as it defines the types of connections between quantum computing nodes, leading to structures such as \acp*{qlan}, quantum network devices or the quantum internet. On the other hand, there is the \textbf{quantum communication channel}, which, as mentioned above, makes use of one of the features of quantum mechanics: qubit entanglement. Finally, the \textbf{communication protocols} are essential for the correct operation of the quantum communication channel, as they define the basic blocks for exchanging quantum information between nodes.

\begin{figure*}[t!]
    \centering
    \includegraphics[width=0.6\textwidth]{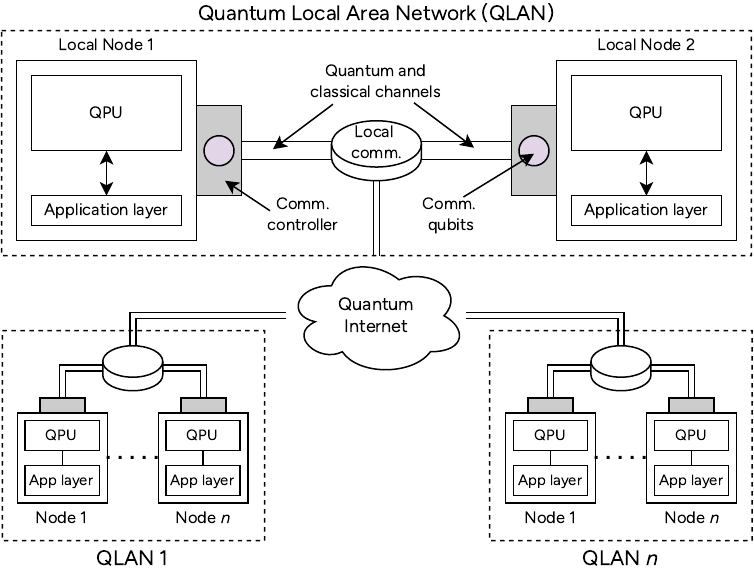}
    \caption{Complex quantum network architecture interconnecting quantum computing nodes of different \acp*{qlan} via the Quantum Internet.}
    \label{fig:sec2:sub-QIN:complex_quantum_network}
\end{figure*}

The \emph{quantum interconnection network} abstracts the communication between different quantum computing nodes. This network can be composed of different types of connections, such as quantum network devices, \acp*{qlan}, or the Quantum Internet. Figure~\ref{fig:sec2:sub-QIN:complex_quantum_network} illustrates a complex example that interconnects quantum computing nodes of different \acp*{qlan} via the Quantum Internet. This network architecture comprises several \acp*{qlan} interconnected through the Quantum Internet, allowing quantum computing nodes to communicate with each other.

The \emph{quantum communication channel}, as its name indicates, represents the abstraction of the quantum channel responsible of the connection between two quantum nodes.  It enables the exchange of quantum information between quantum computing nodes and exploits the principles of quantum mechanics, particularly qubit entanglement. Figure~\ref{fig:sec2:sub-QIN:complex_quantum_network} shows the quantum channel next to a classical channel, which allows operations such as state teleport --an operation that requires both a quantum and a classical channel-- to be implemented.

\input{teledata_telegate}

And the last component of the network layer is also its central element: \emph{quantum communication protocols}. They define the fundamental building blocks for exchanging quantum information between quantum computing nodes. Two of the most important communication protocols are \textit{teledata}~\cite{Bennett1993} and \textit{telegate}~\cite{Gottesman1999}. These protocols exploit qubit entanglement to enable the exchange of quantum information. Both techniques utilise an entangled EPR pair, where one qubit of the pair resides on a \gls*{qpu} and the other is located on a physically separated \gls*{qpu}. These EPR pairs create a link between the two \glspl*{qpu}, allowing quantum data to travel from one \gls*{qpu} to the other by exchanging classical information resulting from measurements of specific qubits. While this work focuses on \texttt{teledata} and \texttt{telegate}, other communication protocols are also available in the literature~\cite{Ferrari2021, Ferrari2023}. Both were selected for this study because they represent two fundamental and widely studied paradigms for distributed quantum communication: quantum state transfer and remote gate application. Their contrasting characteristics make them ideal benchmarks for evaluating abstraction models and compiler decision-making. However, the proposed abstraction model is extensible and can incorporate additional protocols in future work.

Figure \ref{fig:circuits_teledata_telegate} shows the basic structure of the \textit{teledata} (see Figure \ref{subfig:circuits_tele:teledata}) and the \textit{telegate} (see Figure \ref{subfig:circuits_tele:telegate}) protocols. In both techniques, starting from a state $|a\rangle = \alpha|0\rangle + \beta|1\rangle$ in the local \gls*{qpu}\textsubscript{1}, it is necessary that the remote \gls*{qpu}\textsubscript{2} can compute using this information via an EPR pair $|\Phi^+\rangle$. Each protocol is elaborated below:

\begin{itemize}[leftmargin=*]
    \item\textbf{Teledata} protocol transmits the state of the qubit $|a\rangle$ in \gls*{qpu}\textsubscript{1} to an empty qubit in \gls*{qpu}\textsubscript{2}. This transmission involves teleportation of the quantum state, causing the original qubit to collapse upon measurement and transferring its state to the destination qubit.
    
    \item\textbf{Telegate} protocol generates a pair in the state $\alpha|00\rangle + \beta|11\rangle$, where the first qubit is in \gls*{qpu}\textsubscript{1} and the second qubit is in \gls*{qpu}\textsubscript{2}. The second qubit is used as a control qubit for a controlled operation. Considering that the control qubit is in the state $|a\rangle = \alpha|0\rangle + \beta|1\rangle$, using the second qubit of the pair achieves the same effect as performing a controlled operation in \gls*{qpu}\textsubscript{2} with the state of the qubit in \gls*{qpu}\textsubscript{1}.
\end{itemize}

The main difference between teledata and telegate is that in teledata, the state is transferred, and computation is performed locally at the receiving \gls*{qpu}, whereas in telegate, the state is not transferred; instead, quantum gates are controlled remotely. Table~\ref{tab:characteristics_teledata_telegate} compares both techniques by evaluating four key characteristics. It is important to understand the difference between performing an operation "Locally" and "Remotely". A \textbf{local operation} does not require the use of either quantum or classical communications. On the other hand, a \textbf{remote operation} involves the use of quantum or classical communications with other QPUs.

\begin{enumerate}
    \item \emph{Collapsed qubit}: indicates whether the source qubit collapses once the protocol is executed, requiring a qubit reset.
    \item \emph{Entanglement result}: refers to the scope affected by the entanglement generated between the remote and local qubits. This entanglement can be local to the computation node or global to the distributed system.
    \item \emph{Measurements}: describes how the measurements are performed to implement the protocol.
    \item \emph{Number of synchronizations}: the number of synchronizations between the \glspl*{qpu} required to execute the communication protocol.
\end{enumerate}

\input{communication_protocols}

As observed, in the teledata protocol, the qubit collapses when sending the information, necessitating a reset of the qubit afterwards. This occurs because the quantum state is entirely transferred to the target node; thus, operations are performed locally at the destination, and the resulting entanglement is local to the target \gls*{qpu}. Additionally, measurements are performed simultaneously on two qubits local to the \gls*{qpu}\textsubscript{1}, requiring only a single synchronisation between the two \glspl*{qpu}.

In contrast, in the telegate protocol, the quantum information is shared as a reference without measuring the original qubit, eliminating the need to reset it. Sharing a reference implies that the generated entanglement is global to the distributed system --this means that qubits from different QPUs have been entangled--. Furthermore, an initial measurement is performed at the QPU\textsubscript{1}, and a final measurement is conducted on the remote \gls*{qpu}\textsubscript{2}, requiring two separate synchronisations between the \glspl*{qpu}, known as \emph{Cat-Entangler} and \emph{Cat-DisEntangler} (see Figure \ref{subfig:circuits_tele:telegate}). The compiler determines the timing of the second synchronisation (\emph{Cat-DisEnt}), especially when the qubit is no longer in use.

Both protocols have advantages and disadvantages, and there is no clearly superior option. The choice between them depends on the problem to be solved; therefore, the specification of a layered abstraction model will allow the compilation tools to be developed to make an informed decision.

\subsection{Development layer}\label{subsec:development_layer}

In this subsection, the development layer is introduced, designed to provide users with the necessary instructions to work with \ac*{dqc} algorithms while abstracting away the complexities of the network layer. Specifically, two key components for the development layer, shown in Figure~\ref{fig:sec2:layered_abstraction_model}: the \emph{Data Structure for Logical Topology} and the \emph{High-Level Quantum Communication Instructions}. The first component aims to abstract the Quantum Interconnection Network and part of the Quantum Channel by introducing a logical topology data structure that simplifies the development of distributed quantum programs. This logical topology is designed to expose only the number and identity of available quantum processing units (QPUs) to the user, intentionally omitting intermediate network devices such as routers or switches. The second component focuses on abstracting the Quantum Communication Protocols and, to some extent, the Quantum Channel as well, by hiding the implementation details of how communication qubits are generated, through high-level quantum communication instructions.

\begin{figure*}[t!]
    \centering
    \begin{subfigure}[b]{0.4\textwidth}
        \centering
        \includegraphics[width=0.65\textwidth]{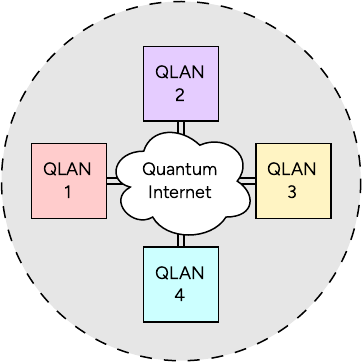}
        \caption{Quantum interconnection network.}
        \label{subfig:sec2:sub-QIN:QIN}
    \end{subfigure}
    \begin{subfigure}[b]{0.4\textwidth}
        \centering
        \includegraphics[width=0.65\textwidth]{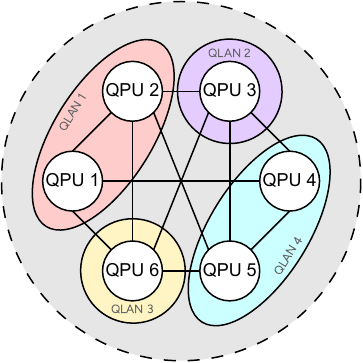}
        \caption{Logical topology.}
        \label{subfig:sec2:sub-QIN:logical_topology}
    \end{subfigure}
    \caption{Comparison between a quantum interconnection network specification and a logical topology for abstract development of distributed quantum programs.}
    \label{fig:sec2:sub-QIN:difference_logical_physical}
\end{figure*}

\subsubsection{Data structure for logical topology}
\label{sec:netqir-data-structure}
In this context, a \textit{process} represents a logical execution entity associated with a QPU, which participates in the distributed computation. Inspired by the classical HPC model, such as MPI, each process in the abstract model is assigned a rank and can be organized into groups and communicators. This abstraction enables coordination of distributed tasks independently of the physical network configuration.

Figure~\ref{fig:sec2:sub-QIN:difference_logical_physical} illustrates the distinction between a physical quantum interconnection network and a logical topology. The Figure~\ref{subfig:sec2:sub-QIN:QIN} shows a simplification of a physical quantum communications infrastructure, where QPUs communicate within QLANs and QLANs communicate with each other over the Quantum Internet. While these components are essential for network functionality, they introduce complexities that are not directly relevant to the programmer. Figure~\ref{subfig:sec2:sub-QIN:logical_topology} presents a logical connection structure to simplify the development of distributed quantum programs. One way to obtain this abstraction is to use data structures such as "Communicator" or "Group" that allow to specify a set of QPUs connected by a logical interconnection. This abstraction allows programmers to focus on high-level application development without being concerned with the underlying network details. The network layer then manages and optimises the actual connections within the network, providing an effective interface between the development layer and the physical infrastructure.

\subsubsection{High-level quantum communication instructions}

High-level directives conform the last piece of abstraction of the development layer. With a clearly defined semantic behaviour, they are able to abstract the underlying communication protocols in \ac*{dqc}. This approach has two main advantages: first, it allows users to develop distributed quantum algorithms without being burdened by the complexity of the communication process; second, it will enable the compiler to understand the precise semantic intent of each function, facilitating the optimisation and the selection of the most appropriate communication protocol. These instructions should prioritize fundamental computational operations, such as data transmission, reception and collective processing, rather than exposing lower-level physical or network mechanisms like entanglement generation, which should remain transparent to the user.

As an example of how a lack of abstraction can negatively impact both performance and software quality, consider the \texttt{entSwap} instruction defined by InQuIR. This instruction explicitly specifies the entanglement swapping procedure, which enables the connection of two quantum nodes that are not directly linked. However, this is fundamentally a low-level problem, as the development layer should not have to manage the connectivity of quantum nodes. Exposing this detail to the development layer might lead users to invoke \texttt{entSwap} unnecessarily, resulting in inefficient calls. Allowing the network---and even lower-level layers---to handle connectivity issues would contribute to more robust software, as these unneeded calls would not be performed.

%% file: teledata_telegate.tex
\begin{figure*}[t!]
    \centering
    \begin{subfigure}[b]{0.45\linewidth}
        \centering
        \begin{adjustbox}{scale=0.8}
        \begin{quantikz}
            \lstick[2]{$\text{QPU}_1$} && \lstick{$|a\rangle$}& \ctrl{1}\gategroup[wires=3, steps=4, style={dashed, rounded corners, fill=green!10, inner sep=2pt}, background]{Teleport}                                                                                           & \gate{H} & \meter{}          & \cwbend{2} & \\
                                       && \lstick[2]{$|\Phi^+\rangle$}   & \targ{}  & \qw      & \meter{}          &            & \\
            \lstick[4]{$\text{QPU}_2$} &&                                & \qw      & \qw      & \gate{X} \vcw{-1} & \gate{Z}   & \swap{1} & \qw        & \qw        & \qw \\
                                       && \lstick{$|0\rangle$}           & \qw      & \qw      & \qw               & \qw        & \targX{} & \ctrl{1} \gategroup[wires=3, steps=2, style={dashed, rounded corners, fill=red!10, inner sep=2pt}, background, label style={label position=below,anchor=north,yshift=-0.2cm}]{Local CZs}   & \ctrl{2}   & \qw \\
                                       && \lstick{$|t_1\rangle$}         & \qw      & \qw      & \qw               & \qw        & \qw      & \control{} & \qw        & \qw \\
                                       && \lstick{$|t_2\rangle$}         & \qw      & \qw      & \qw               & \qw        & \qw      & \qw        & \control{} & \qw 
        \end{quantikz}
        \end{adjustbox}
        \caption{Teledata.}
        \label{subfig:circuits_tele:teledata}
    \end{subfigure}
    %\hspace{0.5cm}
    \begin{subfigure}[b]{0.45\linewidth}
        \centering
         \begin{adjustbox}{scale=0.8}
         \begin{quantikz}
            \lstick[2]{$\text{QPU}_1$} && \lstick{$|a\rangle$}           & \ctrl{1}\gategroup[wires=3, steps=2, style={dashed, rounded corners, fill=green!10, inner sep=2pt}, background]{Cat-Ent}                                                                                & \qw               & \qw        & \qw        & \qw \gategroup[wires=3, steps=2, style={dashed, rounded corners, fill=yellow!10, inner sep=2pt}, background]{Cat-DisEnt}     & \gate{Z} \vcw{2} & \qw \\
                                       && \lstick[2]{$|\Phi^+\rangle$}   & \targ{}  & \meter{}          &            & \\
            \lstick[4]{$\text{QPU}_2$} &&                                & \qw      & \gate{X} \vcw{-1} & \ctrl{1} \gategroup[wires=3, steps=2, style={dashed, rounded corners, fill=red!10, inner                                                                                sep=2pt}, background,label style={label position=below,anchor=north,yshift=-0.2cm}]{Local CZs}   & \ctrl{2}   & \gate{H} & \meter{}         & \qw \\
                                       && \lstick{$|t_1\rangle$}           & \qw      & \qw               & \control{} & \qw        & \qw      & \qw              & \qw \\
                                       && \lstick{$|t_2\rangle$}         & \qw      & \qw      & \qw                 & \control{} & \qw      & \qw              & \qw 
        \end{quantikz}
        \end{adjustbox}
        \caption{Telegate.}
         \label{subfig:circuits_tele:telegate}
    \end{subfigure}
    \caption{Examples of teledata and telegate circuits for the application of CZs.}
    \label{fig:circuits_teledata_telegate}
\end{figure*}

%% file: communication_protocols.tex
\begin{table}[t!]
\centering
\resizebox{\columnwidth}{!}{%
\begin{tabular}{@{}lcccc@{}}
\toprule
\multicolumn{1}{c}{\textbf{Protocol}} & \textbf{\begin{tabular}[c]{@{}c@{}}Collapsed \\ qubit\end{tabular}} & \textbf{\begin{tabular}[c]{@{}c@{}}Entangl.\\ result\end{tabular}} & \textbf{Measures} & \textbf{\begin{tabular}[c]{@{}c@{}}Number\\ syncs\end{tabular}} \\ \midrule
\textbf{Teledata}                            & Yes                                                               & Local                                                                  & Local - Local     & 1                                                                             \\
\textbf{Telegate}                            & No                                                               & Global                                                                 & Local - Remote    & 2                                                                             \\ \bottomrule
\end{tabular}%
}
\caption{Comparative features between teledata and telegate techniques.}
\label{tab:characteristics_teledata_telegate}
\end{table}

%% file: 3_NetQIR.tex
\section{NetQIR: a Quantum Intermediate Representation for Distributed Quantum Computing}
\label{sec:netqir}

This section introduces the \gls*{ir} proposed in this paper: NetQIR, an extension of QIR for \gls*{dqc}. NetQIR is defined according to the layered abstraction model presented in Section~\ref{sec:layered_abstraction_model}, which aims to ensure a certain level of abstraction for the development of efficient and effective tools. It is important to emphasise that an \gls*{ir} is fundamentally a formal specification intended for future developers. To that end, a specification has been created and a Python \gls*{sdk}  developed to test and work with it. This approach allows users to fully understand the specification by experimenting with actual code, thereby facilitating the production of software that employs NetQIR as an \gls*{ir}. Moreover, a grammar has also been developed in ANTLR to allow programmers to translate NetQIR code to specific backends, such as simulators or real systems.

\subsection{NetQIR Specification}
\label{sec4:subsec:netqir_specification}  

This subsection details the NetQIR specification. In doing so, NetQIR defines both data structures and functions. The data structures include two components: \texttt{\%Comm} and \texttt{\%Group}, which correspond to the Communicator and Group described in Section~\ref{sec:netqir-data-structure}, respectively. Regarding functions, NetQIR specifies state functions (which refer not to the quantum state but to the NetQIR program state), data structure functions, and communication functions—the core focus of this work. In addition to this document, the authors provide a more comprehensive specification and detailed documentation on GitHub~\cite{netqir-spec}\footnote{\url{https://netqir.github.io/netqir-spec/}}.

%The specification is structured as follows: Section~\ref{sec4:subsubsec:state_functions} introduces the functions for managing the state of the distributed environment. Next, Section~\ref{sec4:subsubsec:operate_datatypes_functions} defines the instructions for working with the new NetQIR data types. Finally, the most critical part of the specification ---communication operations--- is detailed in Section~\ref{sec4:subsubsec:communication_functions}.

\subsubsection{State functions}\label{sec4:subsubsec:state_functions}
State functions serve as breakpoints where the underlying layers of NetQIR’s abstraction can be defined. For example, these functions provide a point where the compiler can determine when to query and establish connections between different quantum or classical devices. NetQIR introduces two state functions inspired by similar solutions in classical distributed computing frameworks, such as MPI. These functions are:

\begin{itemize}
    \item\texttt{\_\_netqir\_\_initialize()}, which initializes the execution environment.
    \item\texttt{\_\_netqir\_\_finalize()}, which terminates the environment.
\end{itemize}

These functions establish a structured workflow for \gls*{dqc}, ensuring proper initialization and finalization of the execution context.

\subsubsection{Operate datatypes functions}
\label{sec4:subsubsec:operate_datatypes_functions}

In order to abstract from the physical topology, as the development layer explained in Section~\ref{sec:layered_abstraction_model} aims, NetQIR needs to implement a logical topology. For this purpose two already mentioned data structures have been added: \texttt{\%Comm} and \texttt{\%Group}. These will allow the organization of the processes in groups and the establishment of logical topologies that the processor will then be able to link with its physical version. In this abstract model, a process refers to a logical unit of execution associated with a QPU, responsible for performing computations and participating in distributed tasks. This concept, inspired by the notion of processes in classical HPC frameworks like MPI, enables the grouping and coordination of distributed quantum operations. Additionally, data type functions are defined to create or modify the described types and to obtain information about their content at runtime.

NetQIR, as it has been spurred along this work, works akin to MPI. Here another example of the similarities arises, because two key variables are associated with the so-called \texttt{comm\_world}: the process \texttt{rank} and the communicator \texttt{size}. Consequently, both functions will be included:

\begin{itemize}
    \item\texttt{\_\_netqir\_\_comm\_rank}: returns the process \texttt{rank} inside the specified communicator.
    \item\texttt{\_\_netqir\_\_comm\_size}: operation which, from a \texttt{\%Comm} object, returns the number of nodes in that communicator.
\end{itemize}

Moreover, there are also functions established to create, modify or delete \texttt{\%Comm} and \texttt{\%Group}, and, in addition, operations to establish new logical network topologies. For further information on these functions and their use, the reader is referred to the specification~\cite{netqir-spec}.

\subsubsection{Communication functions}\label{sec4:subsubsec:communication_functions}
\input{functions_table}

NetQIR proposes a large set of semantic instructions to improve the construction of DQC algorithms without the need to know the underlying communication protocols. Operations are defined to send and receive classical data, and, within quantum communications, two large sets are created: point-to-point instructions and collective communication routines. These functions are defined in Table~\ref{tab:functions} and explained below.

Collective communication operations are particularly relevant in distributed quantum algorithms, as they enable efficient coordination among multiple QPUs. NetQIR extends classical collective patterns, such as scatter and gather, to the quantum domain, while introducing new abstractions tailored for quantum-specific needs. Among them, the expose operation stands out as a novel contribution. This directive allows multiple QPUs to act upon a shared logical qubit without transferring its state explicitly, leveraging global entanglement to minimize resource consumption. By abstracting the communication protocol and leaving the implementation details to the compiler, expose exploits the layered model's advantages to reduce synchronization overhead and optimize the use of communication qubits in distributed computations.

\paragraph{Point-to-Point communication}
Point-to-point communication in quantum computing parallels that in classical computing, where one node sends or receives information to or from another node. The primary difference is that in the classical case, the information is purely classical, whereas in quantum computing, the information can be classical or quantum. Table~\ref{tab:functions} lists the directives responsible of communication in quantum computing divided into two subgroups: sending and receiving functions. Each sending function corresponds to a receiving one, both of which block the execution of the quantum program. This design ensures that for each send operation at a node there is a corresponding receive operation that unblocks it at the destination node, and vice versa. Mismatches between these could cause an incorrect behavior or compilation errors.

The most basic sending function is \texttt{\_\_netqir\_\_qsend}, representing the part of the circuit on the sending \gls*{qpu}. Additionally, the function \texttt{\_\_netqir\_\_measure\_send} corresponds to sending a classical bit resulting from a measurement, enabling users to develop custom quantum communication protocols. Each of these functions has an \texttt{array} variant for sending or receiving arrays of qubits. It is important to also note that, as it has been said all along the work, the abstraction here performed allows the compiler to choose which communication protocol to use, namely \textit{teledata} or \textit{telegate}. If the user wishes to specify a particular protocol, they can use the specific version of the selected function, for example \texttt{\_\_netqir\_\_qsend\_teledata} in case of wanting to use the \textit{teledata} protocol at sending\footnote{Notably, if a node uses \texttt{\_\_netqir\_\_qsend\_teledata} to send, the receiving node must use \texttt{\_\_netqir\_\_qrecv\_teledata} to receive. Mismatched protocols lead to incorrect behaviour. The general functions \texttt{\_\_netqir\_\_qsend} and \texttt{\_\_netqir\_\_qrecv} offer flexibility by not specifying a protocol, allowing the other node to define it.}.

\paragraph{Collective communication}
While the qubit sending and receiving functions are essential primitives, they may not always be the most efficient choice. Collective communication directives address this by involving multiple \glspl*{qpu} in coordinated operations. They resemble those in classical distributed computing, aiding comprehension for \acs*{hpc} computing users. These functions include \texttt{scatter}, \texttt{gather}, \texttt{reduce} and \texttt{expose}. %Due to the no-cloning theorem, their behaviour differs from that of their classical counterparts. For instance, the \texttt{broadcast} function is replaced by \texttt{expose}, which shares a reference to a qubit instead of copies.

\begin{figure}
    \centering
    \includegraphics[width=0.8\linewidth]{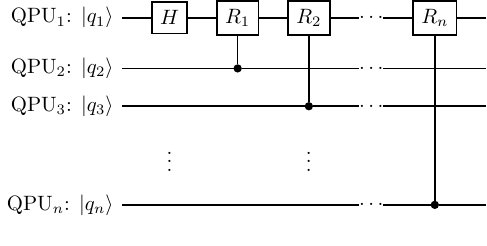}
    \caption{Use case for the \texttt{\_\_netqir\_\_expose} directive on the $|q_1\rangle$ qubit, as it serves as the target for the other remote qubits.}
    \label{fig:sec2:expose}
\end{figure}

\begin{itemize}
    \item \texttt{\_\_netqir\_\_scatter} function distributes an array of qubits from one \gls*{qpu} to several others, enabling parallel processing. 
    \item \texttt{\_\_netqir\_\_gather} function collects qubits from multiple \glspl*{qpu} into a single \gls*{qpu}.
    \item \texttt{\_\_netqir\_\_reduce} directive allows collecting information from multiple remote qubits and applying an operation to obtain a final result. Using \texttt{reduce} simplifies code complexity and enhances computational efficiency compared to sequences of \texttt{qsend} and \texttt{qrecv}.
    \item \texttt{\_\_netqir\_\_expose} directive, which shares a reference to a qubit with other \acp*{qpu}, allowing modifications visible to the entire distributed system. This is particularly useful in operations where all nodes need to use a qubit as a target or control, such as in the distributed \ac*{qft} algorithm~\cite{neumann2020-imperfectQFT}, as illustrated in Figure~\ref{fig:sec2:expose}. In this circuit, a sequence of controlled-phase gates is applied between one target qubit and several control qubits located in different \acp*{qpu}. Using \texttt{expose}, the target qubit can be made available to all control units without physically transferring its state, allowing each QPU to apply its operation as if the qubit were local. This significantly reduces the communication overhead in such distributed scenarios. Additionally, Figure~\ref{fig:sec2:expose:implementation} shows a possible implementation of the expose operation using a GHZ state to connect the \acp*{qpu}, achieving the state $\alpha|00\ldots00\rangle + \beta|11\ldots11\rangle$ with the exposed state being $\alpha|0\rangle + \beta|1\rangle$. This implementation is not part of the development layer but represents one of the strategies available to the compiler could choose depending on the underlying physically network.
\end{itemize}

\begin{figure}
    \centering
    \includegraphics[width=0.9\linewidth]{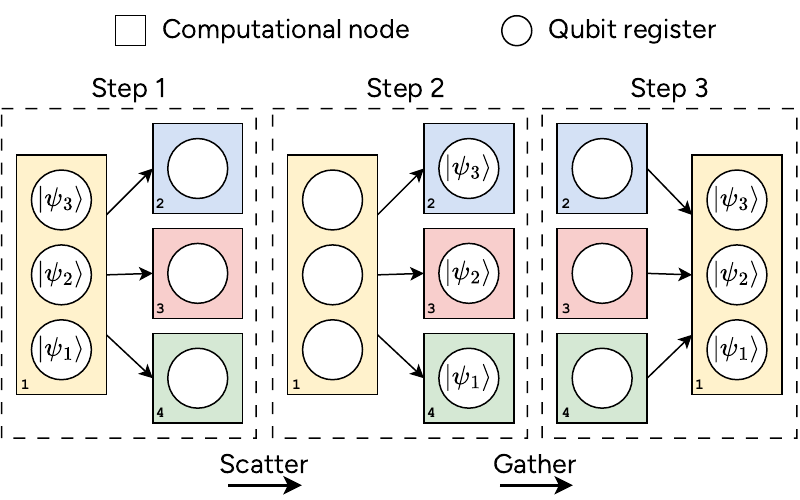}
    \caption{An example of using a scatter teledata operation on a qubit array (steps 1 to 2) and the inverse gather teledata operation (steps 2 to 3) between 4 QPUs (labeled in each square).}
    \label{fig:scatter_gather}
\end{figure}

Similar to point-to-point functions, collective directives have \textit{teledata} and \textit{telegate} variants. Figure~\ref{fig:scatter_gather} illustrates the use of scatter and gather operations using the teledata protocol.

\begin{figure}
    \centering
    \resizebox{0.9\linewidth}{!}{
    \input{expose}
    }
    \caption{Possible implementation of the \texttt{\_\_netqir\_\_expose} directive on the qubit $|\psi\rangle_{QPU_1}$, which is the target of the rest of the remote qubits. Distributed operations would be performed between \texttt{cat-ent} and \texttt{cat-dis}.}
    \label{fig:sec2:expose:implementation}
\end{figure}

\subsection{NetQIR SDK: PyNetQIR}\label{sec4:subsec:netqir_sdk}
The NetQIR specification constitutes the central core of an \gls*{ir}, serving as the common starting point for developers in the field of distributed quantum computing. To support developers in building distributed quantum applications, the NetQIR specification is accompanied by an open-source Python \gls*{sdk} designed to facilitate the generation of NetQIR code in Python-based environments. This tool, named PyNetQIR, is available in the project's GitHub repository\footnote{\url{https://github.com/netqir/netqir-sdk}}.

This SDK enables the generation of NetQIR code from high-level Python scripts, following a structured execution model composed of three main components: \textit{Operations}, \textit{Scopes}, and \textit{Executors}. This is provided as a summary, for more information please consult the open source code repository.

\begin{figure*}[t!]
    \centering
    \begin{subfigure}[b]{0.5\textwidth}
        \centering
        \includegraphics[width=\textwidth]{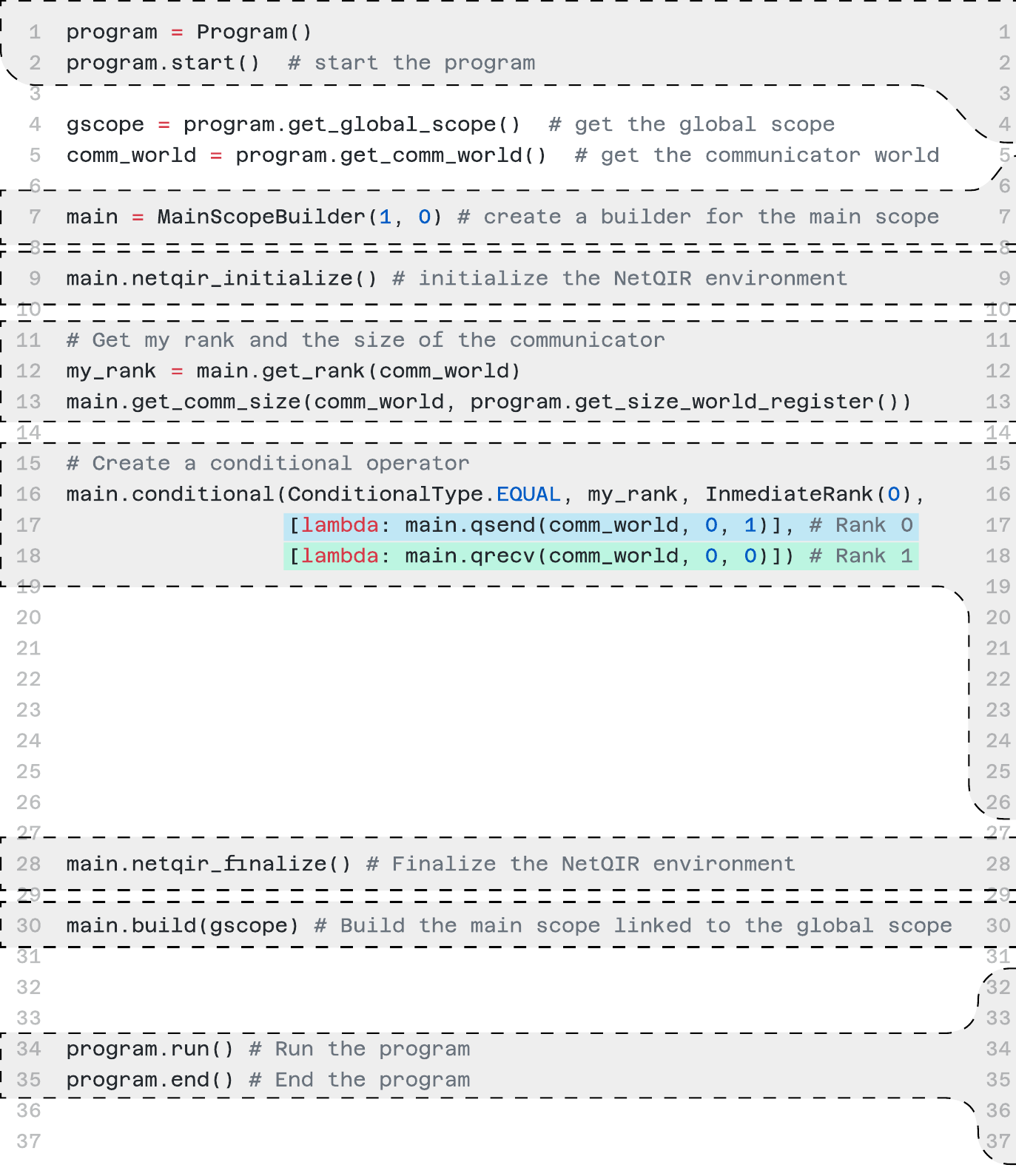}
        \caption{PyNetQIR code.}
        \label{code:pynetqir}
    \end{subfigure}%
    \begin{subfigure}[b]{0.5\textwidth}
        \centering
        \includegraphics[width=\textwidth]{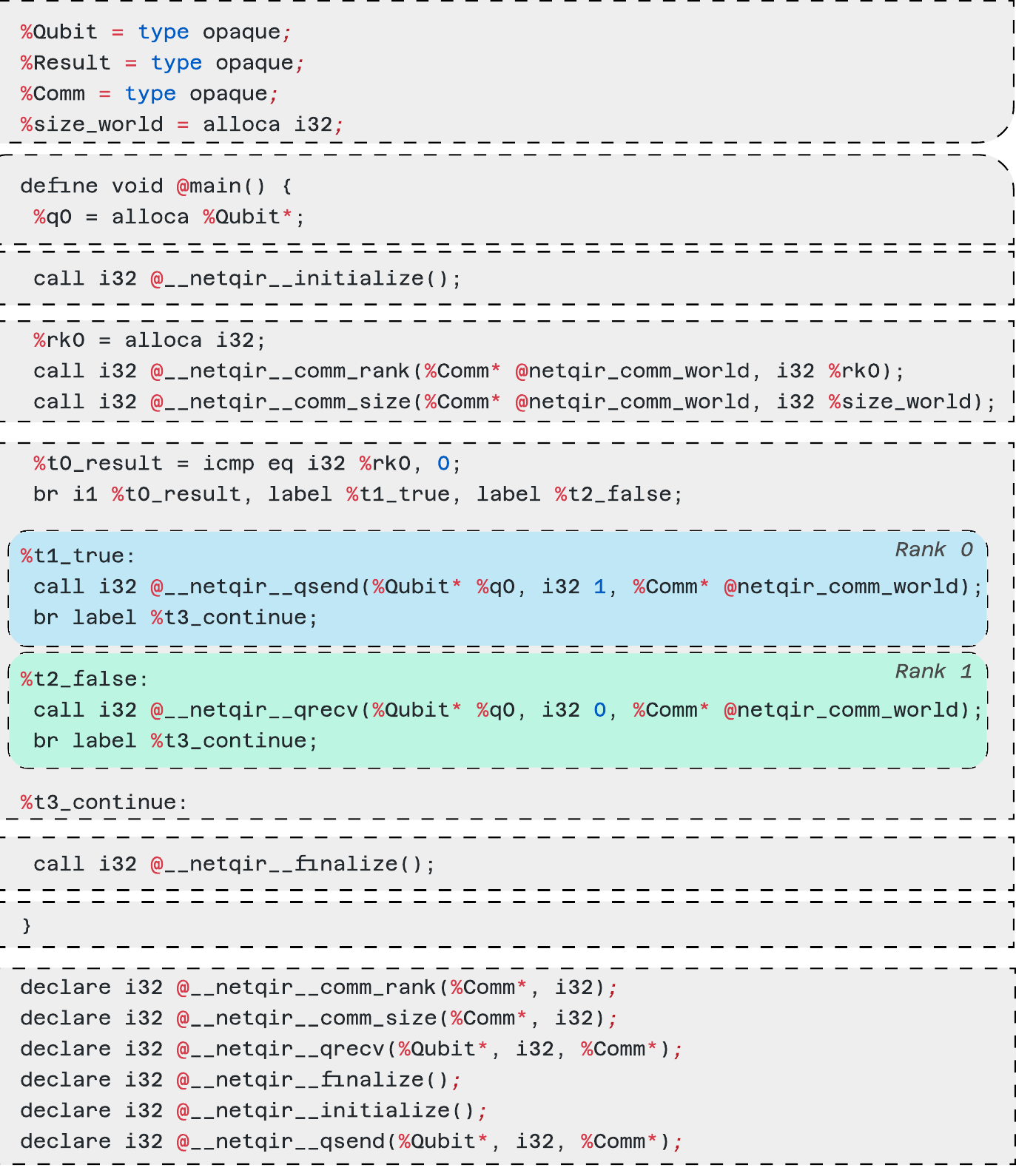}
        \caption{Generated NetQIR from PyNetQIR code (reduced for document legibility).}
        \label{code:pynetqir-to-netqir}
    \end{subfigure}%
    \caption{Generation of NetQIR code of teleportation circuit using PyNetQIR.} 
    \label{code:chap4:generation_pynetqir}
\end{figure*}

\begin{itemize}
    \item\textbf{Operations} represent the basic actions of a NetQIR program. These may include quantum instructions (such as gate applications or qubit transmissions) as well as classical instructions (like conditionals or communicator queries), allowing for hybrid programming. This hybrid approach extends the LLVM framework to integrate quantum semantics while maintaining compatibility with classical logic.

    \item\textbf{Scopes} are hierarchical structures that group operations logically, similar to blocks in traditional programming. The SDK provides builders (e.g., \texttt{MainScopeBuilder}) to construct these scopes and manage their contents efficiently.

    \item\textbf{Executors} are responsible for interpreting or compiling the operations within a scope. In the example shown in Figure~\ref{code:chap4:generation_pynetqir}, a \texttt{PrinterExecutor} is used to emit the resulting NetQIR code, although other executors could target simulators or hardware backends in future implementations.
\end{itemize}

The typical flow of a NetQIR program using this SDK begins by initializing the program and obtaining the global scope and communicator. Figure~\ref{code:chap4:generation_pynetqir} shows an example of this process, where a quantum state is transferred from one \ac{qpu} to another using the \texttt{qsend} and \texttt{qrecv} directives. Within the main scope, the NetQIR environment is initialized and the rank and size of the communicator are retrieved. By leveraging ranks and communicators, the SDK mirrors the structure of classical distributed frameworks like MPI, allowing developers to adopt familiar patterns when building quantum programs.

A conditional operator based on the process rank is defined: if the rank is \texttt{0}, the process performs a quantum send (\texttt{qsend}); if it is \texttt{1}, it performs a quantum receive (\texttt{qrecv})\footnote{Several examples, including this one, are available in the repository: \url{https://github.com/NetQIR/netqir-sdk/tree/main/python/examples}}. The environment is then finalized and the program is executed using the Executor. This example demonstrates how NetQIR supports modular and semantically clear construction of distributed quantum applications using a model inspired by classical distributed computing. 

% \begin{figure*}[t!]
%     \centering
%     \begin{subfigure}[b]{0.39\textwidth}
%         \centering
%         \lstinputlisting[language=python,numbers=left]{code/example_pynetqir.py}
%         \caption{PyNetQIR code.}
%         \label{code:pynetqir}
%     \end{subfigure}
%     \hfill
%     \begin{subfigure}[b]{0.59\textwidth}
%         \centering
%         \lstinputlisting[numbers=left]{code/result_example_pynetqir_netqir.ll}
%         \caption{Generated NetQIR from PyNetQIR code (reduced for document legibility).}
%         \label{code:pynetqir-to-netqir}
%     \end{subfigure}
%     \caption{Generation of NetQIR code of teleportation circuit using PyNetQIR.} 
%     \label{code:chap4:generation_pynetqir}
% \end{figure*}

\subsection{NetQIR ANTLR grammar}\label{sec4:subsec:netqir_grammar}
Once the NetQIR specification has been defined and an \gls*{sdk}  for translating Python code into NetQIR has been implemented, it becomes essential to provide future developers with a tool for extending NetQIR. This includes developing new high-level languages that compile to NetQIR and translating NetQIR into low-level machine instructions for quantum devices. To facilitate this, a formal grammar definition is introduced, leveraging the ANTLR~\cite{Parr1995antlr, parr2013antlr-definitive} specification to enable structured parsing and transformation of NetQIR code. The primary advantage of defining this grammar is that future developers can choose the target programming language for their NetQIR parser. ANTLR provides automatic code generation from its grammar definitions, supporting a wide range of well-known high-level programming languages. This flexibility facilitates the integration of NetQIR into diverse software ecosystems, enabling seamless adoption across different development environments. The grammar defined for NetQIR is encoded in the GitHub repository \texttt{netqir-grammar}\footnote{NetQIR grammar: \url{https://github.com/NetQIR/netqir-grammar/}}. It is important to note that this grammar aims to classify the different categories of NetQIR functions specified earlier. This approach allows developers to generate more specialized listeners for the generated AST tree, providing, for instance, information on whether they are programming a \texttt{qsend} function with modifiers such as \texttt{teledata} or \texttt{array}.

\begin{figure*}[t!]
    \centering
    \includegraphics[width=0.7\linewidth]{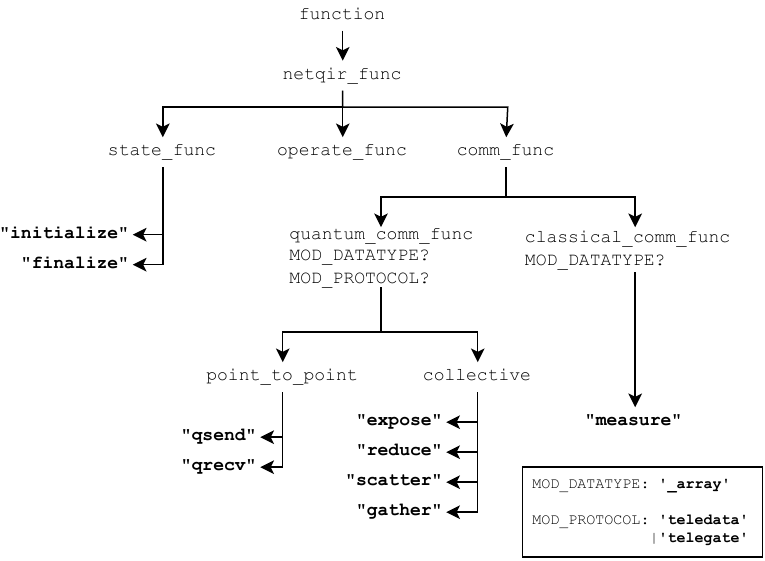}
    \caption{NetQIR Grammar for the classification of functions in its specification.}
    \label{fig:sec4:netqir-grammar}
\end{figure*}

Figure~\ref{fig:sec4:netqir-grammar} presents a syntax tree illustrating the structure of the defined grammar, where bold elements represent lexical tokens. It can be observed that both quantum and classical communication functions support modifiers, such as ``\texttt{array}" for sending quantum or classical arrays, and protocol specifications like ``\texttt{telegate}" or ``\texttt{teledata}". It is important to note that both modifiers are optional (hence the use of the \texttt{?} symbol). If no protocol is explicitly specified, the compiler will automatically select the most optimal one based on the execution context.

%Finally, it is crucial to understand that this grammar functions as an extension and enhancement of the existing LLVM grammar, which is widely recognized and well-established in the field of classical compiler design. By building upon LLVM, we leverage a vast ecosystem of pre-existing tools and optimizations, ensuring a seamless integration for hybrid quantum-classical programming.

%% file: functions_table.tex
\begin{table*}[t!]
\centering
\resizebox{\textwidth}{!}{%
\begin{tabular}{@{}llll@{}}
\toprule
\multicolumn{4}{c}{\textbf{Point-to-point communication functions}}                                                                                           \\ \midrule
\multicolumn{2}{c}{\textbf{Sending functions}}                                & \multicolumn{2}{c}{\textbf{Receiving functions}}                              \\ \midrule
\_\_netqir\_\_qsend\_array           & (Array*, i32, i32, Comm*)              & \_\_netqir\_\_qrecv\_array           & (Array**, i32, i32, Comm*)             \\
\_\_netqir\_\_qsend\_array\_teledata & (Array*, i32, i32, Comm*)              & \_\_netqir\_\_qrecv\_array\_teledata & (Array**, i32, i32, Comm*)             \\
\_\_netqir\_\_qsend\_array\_telegate & (Array*, i32, i32, Comm*)              & \_\_netqir\_\_qrecv\_array\_telegate & (Array**, i32, i32, Comm*)             \\
\_\_netqir\_\_qsend                  & (Qubit*, i32, Comm*)                   & \_\_netqir\_\_qrecv                  & (Qubit**, i32, Comm*)                  \\
\_\_netqir\_\_qsend\_teledata        & (Qubit*, i32, Comm*)                   & \_\_netqir\_\_qrecv\_teledata        & (Qubit**, i32, Comm*)                  \\
\_\_netqir\_\_qsend\_telegate        & (Qubit*, i32, Comm*)                   & \_\_netqir\_\_qrecv\_telegate        & (Qubit**, i32, Comm*)                  \\
\_\_netqir\_\_measure\_send\_array   & (Array*, i32, i32, Comm*)              & \_\_netqir\_\_measure\_recv\_array   & (i1*, i32, i32, Comm*)                 \\
\_\_netqir\_\_measure\_send          & (Qubit*, i32, Comm*)                   & \_\_netqir\_\_measure\_recv          & (i1*, i32, i32, Comm*)                 \\ \midrule
\multicolumn{1}{c}{}                 & \multicolumn{1}{c}{}                   &                                      &                                        \\ \midrule
\multicolumn{4}{c}{\textbf{Collective communication functions}}                                                                                               \\ \midrule
\_\_netqir\_\_scatter                & (Array*, i32, Array*, i32, i32, Comm*) & \_\_netqir\_\_expose                 & (Qubit*, i32, Comm*)                   \\
\_\_netqir\_\_scatter\_teledata      & (Array*, i32, Array*, i32, i32, Comm*) & \_\_netqir\_\_expose\_array          & (Array*, i32, i32, Comm*)              \\
\_\_netqir\_\_scatter\_telegate      & (Array*, i32, Array*, i32, i32, Comm*) &                                      &                                        \\
\_\_netqir\_\_gather                 & (Array*, i32, Array*, i32, i32, Comm*) & \_\_netqir\_\_reduce                 & (Array*, i32, Array*, i32, i32, Comm*) \\
\_\_netqir\_\_gather\_teledata       & (Array*, i32, Array*, i32, i32, Comm*) & \_\_netqir\_\_reduce\_teledata       & (Array*, i32, Array*, i32, i32, Comm*) \\
\_\_netqir\_\_gather\_telegate       & (Array*, i32, Array*, i32, i32, Comm*) & \_\_netqir\_\_reduce\_telegate       & (Array*, i32, Array*, i32, i32, Comm*) \\ \bottomrule
\end{tabular}%
}
\caption{NetQIR functions: point-to-point and collective.}
\label{tab:functions}
\end{table*}

%% file: expose.tex
\begin{quantikz}
    \lstick[1]{\ket{\psi}$_{QPU_1}$} & \ctrl{1} & \qw & \qw & \qw & \qw & \gate{Z} & \\
    \lstick[5]{\ket{GHZ}} QPU_1 & \targ{} & \meter{} \\
    QPU_2  & \qw      & \gate{X} \vcw{-1} & \qw \ \ldots\ & \gate{H} & \meter{} & \cwbend{-2} & \\
    & \vdots    &                   & \vdots              &          &   \vdots       &             & \\
    QPU_{n-1}  & \qw      & \gate{X} \vcw{-2} & \qw \ \ldots\ & \gate{H} & \meter{} & \cwbend{-4} & \\
    QPU_n & \qw      & \gate{X} \vcw{-1} & \qw \ \ldots\ & \gate{H} & \meter{} & \cwbend{-5} & \\
\end{quantikz}

%% file: 4_Results.tex
\section{Results}\label{sec:results}
This section presents two complementary analyses to support the design rationale behind NetQIR and to evaluate its relevance for distributed quantum computing.

First, a \textbf{comparative analysis with state-of-the-art languages} is performed to assess how existing DQC intermediate representations and frameworks align with the layered abstraction model proposed in this work. This comparison identifies the extent to which each solution abstracts the complexities of distributed quantum execution at the network and development layers.

Second, a \textbf{justification of the layered abstraction model} is provided, with a focus on the benefits introduced by collective communication directives. In particular, this analysis demonstrates how high-level operations, such as \texttt{scatter}, \texttt{gather}, or the novel \texttt{expose}, empower the compiler to select the most appropriate communication strategy depending on the system’s topology or the specific requirements of a given algorithm. Rather than prescribing low-level instructions, these abstractions enable optimization and adaptation to the execution context, ultimately leading to more efficient resource usage and scalability.

Together, these two analyses highlight the practical and conceptual value of adopting a layered abstraction model and the role of NetQIR in bridging high-level algorithm design with efficient distributed quantum execution.

\subsection{Comparison with State-of-the-Art Languages}\label{sec:results:comparison}

%This paper proposes a model of abstraction layers for developing new \ac*{dqc} systems. In addition, a \ac*{ir} is specified that aims to implement the proposed development layer. It is interesting to be able to evaluate the different programming languages designed for \ac*{dqc} that exist in the literature, but taking into account that some of them, such as \ac*{qmpi} or NetQIR, are specifications and lack implementation a qualitative evaluation is made.
This section presents a comparative analysis of existing \ac*{dqc} languages and \acp*{ir}, focusing on their alignment with the layered abstraction model proposed in Section~\ref{sec:layered_abstraction_model}. This model defines essential elements across two layers ---the network layer and the development layer--- designed to facilitate the efficient development of \ac*{dqc} tools. The evaluation criteria include key aspects such as the ones outlined below: \\

\noindent\textbf{Network layer}: the network layer contains the necessary characteristics to define a correct quantum-classical connection between different nodes, abstracting from the underlying physical complexities, like: %characteristics defined in this work.

\begin{itemize}[leftmargin=*]
    \item\textbf{Quantum Channel Abstraction}: evaluates whether the language abstracts the quantum channels used for inter-node communication, essential for managing quantum information transfer. 
    \item\textbf{Quantum Interconnection Network}: assesses the language's ability to abstract the structure of quantum interconnection networks connecting multiple \acp*{qpu}, a foundational aspect for scalable \ac*{qpu} architectures. 
    \item\textbf{Communication Protocols}: identifies whether the abstraction of the communication protocol is allowed or has to be defined by the user when programming. Abstraction is essential to allow the compiler to optimise techniques according to the context.\\
\end{itemize}
    
\noindent\textbf{Development layer}: programming languages for \ac*{dqc} must incorporate features that provides users to perform distributed quantum computing while abstracting away the complexities of the underlying quantum network.
    \begin{itemize}[leftmargin=*]
       \item\textbf{High-Level Quantum Communication Instructions}: considers whether the language provides high-level commands to simplify distributed quantum operations, facilitating programming efficiency and code readability. These instructions allow the compiler to provide functions with semantic context, allowing it to perform optimisations between the different possible physical implementations.
       
       \item \textbf{Data Structures for Logical Topology}: determines the language's support for data structures that abstract the physical topology of the distributed system over a logical topology, allowing easier specification of quantum node relationships and delegating to the compiler the responsibility for matching the code to the target topology.
    \end{itemize}
    
\noindent\textbf{Other characteristics}: key features for an \gls*{ir} to enable the correct and efficient development of new tools.
\begin{itemize}[leftmargin=*]
       \item\textbf{Real Quantum Computing Inspired:} specifies whether the language is intended to perform real quantum computation or only simulated quantum computation. This feature aims to eliminate all languages that include instructions that are not allowed in the quantum model, such as perfect copying of generic states.

       \item\textbf{Hybrid programming:} this feature is crucial for enabling the orchestration between classical and quantum computing devices, facilitating tasks such as optimization problems.
\end{itemize}

Table~\ref{tab:qualitative_comparative} shows the comparison between the different languages discussed in Section~\ref{sec:related-work} (related work) together with the \ac*{ir} proposed in this work: NetQIR. 

\input{qualitative_comparative}
In particular, NetQASM and InQuIR do not fully abstract the quantum channel. InQuIR requires the programmer to explicitly call operations such as \texttt{genEnt} to create entangled pairs and \texttt{entSwp} to perform entanglement swapping between QPUs. This exposes the physical routing of qubits and limits the flexibility of the compiler to adapt to different network configurations. Similarly, NetQASM uses \texttt{create\_epr} to establish entanglement between nodes, providing only a minimal abstraction and restricting the channel model to EPR-based links, without considering other communication resources such as GHZ states or multi-party entanglement.

Regarding the quantum interconnection network, both approaches offer at best a partial abstraction. Although entanglement-based communication is used, the fact that the programmer must manage the flow of entangled pairs (e.g., through manual swapping or addressing specific qubits) means that the logical network structure is not decoupled from the physical implementation. For instance, in InQuIR, if a QPU wants to communicate with a non-adjacent node, the user must explicitly define a chain of entSwp instructions. This introduces rigid dependencies on the physical topology and prevents the compiler from transparently handling the routing of quantum information.

Concerning the communication protocols, there is no abstraction in this feature because the programmer has to decide how to interact with the communication qubits. Furthermore, the development layer is not implemented as neither quantum communication instructions nor structures to define a logical topology are defined. Hybrid programming is not allowed in languages such as NetQASM or InQuIR which are specific to quantum device programming. 

With respect to QMPI, it allows abstraction in the fields indicated, except that it is not intended for real quantum computation because it has collective operations that are not meaningful due to the non-cloning theorem, such as the \texttt{Allscatter} or \texttt{Allgather} operation. It is also important to specify that QMPI is a message passing interface, so it is not an \gls*{ir}, just as MPI is not an \gls*{ir}. On the other hand, QMPI, being a message passing interface and not an \gls*{ir}, could allow this type of programming depending on its future implementations. 

Finally, NetQIR, the intermediate representation for DQC proposed in this paper, would meet the above requirements by abstracting each part of the layered model. NetQIR, by extending QIR, which in turn extends LLVM, ensures hybrid programming by integrating quantum and classical programming in the same \gls*{ir}.

\subsection{Justification of the Layered Abstraction Model}\label{sec:results:layered_model}
This section provides a quantitative justification for the layered abstraction model introduced in this work. The objective is to evaluate the practical advantages of high-level collective operations—enabled by the development layer—in optimizing communication strategies for distributed quantum computing.

The analysis focuses on how resource consumption varies depending on the selected communication protocol, the physical topology of the quantum network, and the number of \acp{qpu} involved. By abstracting these aspects through semantic directives and delegating the decision-making to the compiler, the model allows for adaptation to the underlying infrastructure and algorithmic needs, improving overall efficiency.

Two main metrics are considered: the total number of \textbf{communication qubits consumed} during circuit execution, and the number of \textbf{communication qubits each QPU must reserve} to support the communication process. These results serve to highlight the benefits of decoupling communication details from the algorithmic description, as promoted by the layered abstraction model.

To do this, the circuit in Figure~\ref{circuit:qft} was used, which represents the computation of a \ac*{qft} using $n+1$ qubits. This process can be separated into $n$ epochs, where in epoch $i$ controlled gates are applied to qubit $|x_i\rangle$, equivalent to an \texttt{expose} function. In this use case, the QPU $i$ is assigned the qubit $|x_i\rangle$.

\input{qft}

\begin{figure*}[t!]
    \centering
    \begin{subfigure}[b]{0.40\textwidth}
        \centering
        \includegraphics[width=0.728\textwidth]{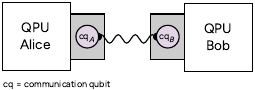}
        \caption{Direct interconnection.}
        \label{subfig:sec3:direct_connection}
    \end{subfigure}
    \hspace{0.5cm}
    \begin{subfigure}[b]{0.40\textwidth}
        \centering
        \includegraphics[width=\textwidth]{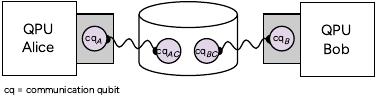}
        \caption{Interconnection via a communicator.}
        \label{subfig:sec3:inter_via_communicator}
    \end{subfigure}
    \caption{Examples of different quantum network topologies for connecting QPUs within a QLAN.}
    \label{fig:sec3:example_topologies}
\end{figure*}

On the other hand, the topologies to be tested are shown in Figure~\ref{fig:sec3:example_topologies}, which are the direct connection option and the interconnection option via a communicator.

\begin{figure*}[t]
    \centering
    \begin{subfigure}[b]{0.31\textwidth}
        \centering
        \resizebox{!}{0.81\textwidth}{\input{q_consumed_direct_connection}}
        \caption{Directly Connected Topology (D-C)}
        \label{subfig:sec4:comm_qubits_consumed:d-c-topology}
    \end{subfigure}
    \hfill
    \begin{subfigure}[b]{0.31\textwidth}
        \centering
        \resizebox{!}{0.81\textwidth}{\input{q_consumed_t-1c}}
        \caption{Topology via one communicator (T1-C)}
        \label{subfig:sec4:comm_qubits_consumed:t-1c-topology}
    \end{subfigure}
    \hfill
    \begin{subfigure}[b]{0.31\textwidth}
        \centering
        \resizebox{!}{0.81\textwidth}{\input{q_comm}}
        \caption{Comm. qubits needed for a QPU.}
        \label{subfig:sec4:comm_qubits_needed}
    \end{subfigure}
    \caption{Comparison of the number of communication qubits consumed and needed for the QFT circuit, in function of the number of connected QPUs, the communication protocol used and the existing physical topology.}
    \label{fig:sec4:comm_qubits_consumed}
\end{figure*}

\begin{itemize}[leftmargin=*]
    \item\textbf{Direct interconnection}: this type of connection is peer-to-peer so that each QPU is connected to each of the other QPUs. This allows a direct connection between each pair of QPUs without unnecessary hops but has the disadvantage of having a tightly coupled network, making it problematic to add new nodes and requiring a large number of communication qubits. 
    \item\textbf{Topology via one communicator}: In this case, access to the distributed system is managed through a quantum communicator (or quantum router). Each \ac{qpu} is connected only to this central node, which is responsible for relaying quantum information between them. One of the main advantages of this abstraction is that it allows the compiler to choose the most suitable strategy depending on the context. In some situations, the communicator may assist the routing process by performing entanglement swapping between QPUs. In other cases, the communicator might directly establish entanglement links with the target QPU and transfer the qubit's state or apply remote operations, acting as an active participant in the communication. This flexibility highlights the benefit of leaving implementation decisions to the lower layers of the stack.
\end{itemize}

Continuing with the communication protocols evaluated, the analysis considers the use of \texttt{teledata}, \texttt{telegate}, and \texttt{expose} in both topologies. For the \texttt{expose} function, the implementation shown in Figure~\ref{fig:sec2:expose:implementation} is used, where a GHZ state is generated to connect all involved \acp{qpu}.

In the case of \texttt{teledata}, the source QPU must send the qubit to the destination node, allow the operation to be performed, and then retrieve the updated state—requiring two EPR pairs per operation. The \texttt{telegate} protocol, which executes the operation remotely without moving the qubit, requires only a single EPR pair. Lastly, the \texttt{expose} directive leverages a shared GHZ state, allowing multiple QPUs to access the same logical qubit. This significantly reduces the number of required communication qubits, especially as the number of QPUs increases, since the cost of GHZ generation is shared across the participants.

As can be seen in Figure~\ref{subfig:sec4:comm_qubits_consumed:d-c-topology} and \ref{subfig:sec4:comm_qubits_consumed:t-1c-topology}, for both topologies, the protocol with the highest qubit consumption is \texttt{teledata}, followed by \texttt{telegate}, and finally \texttt{expose}, which shows the lowest consumption. This highlights the importance of selecting an appropriate communication protocol for each specific task. By using high-level semantic functions, the compiler gains the flexibility to optimise communication by selecting the most efficient strategy based on the topology and algorithmic context.

On the other hand, Figure~\ref{subfig:sec4:comm_qubits_needed} shows the number of communication qubits each QPU must reserve to communicate with the distributed system. As expected, this number grows linearly in the case of a directly connected topology, requiring one or two additional communication qubits for each new QPU added to the system. In contrast, in the via one communicator topology, this requirement remains constant, as only one or two communication qubits are needed per QPU, regardless of the system size, since the quantum communicator handles the interconnections.

Overall, these results reinforce the value of the proposed layered abstraction model. By abstracting protocol and topology details through collective operations, the compiler is empowered to make optimal decisions, improving scalability and resource efficiency in distributed quantum systems.

% \begin{figure}
%     \centering
%     \input{plots/q_comm}
%     \caption{number of communication qubits that a single QPU in the QFT circuit needs to connect to the rest of the distributed system.}
%     \label{fig:sec4:comm_qubits_needed}
% \end{figure}

%% file: qualitative_comparative.tex
\begin{table*}[t]
\resizebox{\textwidth}{!}{%
\begin{tabular}{@{}lccccccc@{}}
\toprule
\multicolumn{1}{c}{\multirow{2}{*}{\textbf{Language}}} & \multicolumn{3}{c}{\textbf{Network layer}}                                                                                                                                                                                       & \multicolumn{2}{c}{\textbf{Development layer}}                                                                                                                                    & \multicolumn{2}{c}{\textbf{Other characteristics}}                                                                                                         \\ \cmidrule(l){2-8} 
\multicolumn{1}{c}{}                                   & \textbf{\begin{tabular}[c]{@{}c@{}}Quantum \\ channel\end{tabular}} & \textbf{\begin{tabular}[c]{@{}c@{}}Q. Interconnection\\ Network\end{tabular}} & \textbf{\begin{tabular}[c]{@{}c@{}}Communication\\ protocols\end{tabular}} & \textbf{\begin{tabular}[c]{@{}c@{}}High-level quantum\\ comm. instructions\end{tabular}} & \textbf{\begin{tabular}[c]{@{}c@{}}Data structure for\\ logical topology\end{tabular}} & \textbf{\begin{tabular}[c]{@{}c@{}}Real Quantum\\ Computing inspired\end{tabular}} & \textbf{\begin{tabular}[c]{@{}c@{}}Hybrid\\ programming\end{tabular}} \\ \midrule
NetQASM                                                & $\boldsymbol{\sim}$                                                 & $\boldsymbol{\sim}$                                                           & \xmark                                                                     & \xmark                                                                                   & \xmark                                                                                 & \cmark                                                                             & \xmark                                                                \\
InQuIR                                                 & $\boldsymbol{\sim}$                                                 & \xmark                                                           & \xmark                                                                     & \xmark                                                                                   & \xmark                                                                                 & \cmark                                                                             & \xmark                                                                \\
QMPI                                                   & \cmark                                                              & \cmark                                                                        & \cmark                                                                     & \cmark                                                                                   & \cmark                                                                                 & \xmark                                                                             & $\boldsymbol{\sim}$                                                                \\
NetQIR                                                 & \cmark                                                              & \cmark                                                                        & \cmark                                                                     & \cmark                                                                                   & \cmark                                                                                 & \cmark                                                                             & \cmark                                                                \\ \bottomrule
\end{tabular}%
}
\caption{Qualitative comparison table between the different languages selected for DQC programming.}
\label{tab:qualitative_comparative}
\end{table*}

%% file: qft.tex
\begin{figure}[t!]
    \centering
    \resizebox{0.5\textwidth}{!}{
    \begin{quantikz}
        \lstick{$\ket{x_0}$}        & \gate{H}  & \gate{R_1}        & \qw \ \ldots\     & \gate{R_n}    & \qw           & \qw       & \qw & \qw & \qw & \qw & \qw & \qw & \qw & \qw \\
        \lstick{$\ket{x_1}$}        & \qw        & \ctrl{-1}        & \qw           & \qw           & \qw           & \qw  & \qw           & \qw       & \qw & \qw & \qw & \qw & \qw & \qw\\
        \wave                       &            &                  &               &               &               &           &            &        &  &  &  &  & & \\
        \lstick{$\ket{x_{i}}$}      & \qw        & \qw              & \qw           & \qw           & \gate{H}\gategroup[7,steps=6,style={dashed, inner sep=6pt}, label style={label position=below,anchor=north,yshift=-0.2cm}]{$i^{th}$-{\sc expose}}      & \gate{R_1} & \qw \ \ldots\ & \gate{R_k} & \qw \ \ldots\ & \gate{R_{n-i}}       & \qw & \qw  & \qw & \qw\\
        \lstick{$\ket{x_{i+1}}$}    & \qw        & \qw              & \qw           & \qw           & \qw           & \ctrl{-1}  & \qw           & \qw       & \qw & \qw & \qw & \qw & \qw & \qw\\
                                    & \vdots     &                  &               &               & \vdots        &           &            & \vdots       &  &  &  \vdots  & & &  & \\
        \lstick{$\ket{x_{i+k}}$}    & \qw        & \qw              & \qw           & \qw           & \qw           & \qw  & \qw           & \ctrl{-3}       & \qw & \qw & \qw & \qw & \qw & \qw\\
        \wave                       &            &                  &               &               &               &           &            &        & &  &  &  & &\\
        \lstick{$\ket{x_{n-1}}$}    & \qw        & \qw              & \qw           & \qw           & \qw           & \qw       & \qw           & \qw       & \qw & \qw & \qw & \gate{H}\gategroup[2,steps=2,style={dashed, inner sep=6pt}, label style={label position=below,anchor=north,yshift=-0.2cm}]{$n^{th}$-{\sc expose}} & \gate{R_1} & \qw\\
        \lstick{$\ket{x_n}$}        & \qw        & \qw              & \qw           & \ctrl{-9}     & \qw           & \qw       & \qw           & \qw       & \qw & \ctrl{-6} & \qw & \qw & \ctrl{-1} & \qw \\
                               &            &                  &                   &               &               &               &               &           &            &        &  &  &  &  & & 
    \end{quantikz}
    }
    \caption{Circuit for the calculation of the Quantum Fourier Transformation (QFT), where $n$ calculations are performed on $n$ qubits}
    \label{circuit:qft}
\end{figure}

%% file: q_consumed_direct_connection.tex
\begin{tikzpicture}
        \begin{axis}[
            xlabel={Number of QPUs connected},
            ylabel={Communication Qubits consumed},
            ymin=-5, ymax=260, 
            xmin=2, xmax=11,
            legend style={at={(0.25,0.95)}, anchor=north, draw=none, fill=none},
            legend cell align={left},
            xtick={1,2,3,4,5,6,7,8,9,10}
        ]
        
        % Teledata
        \addplot+[mark=*, mark options={solid}, thick] plot coordinates {
            (2,4) (3,12) (4,24) (5,40) (6,60) (7,84) (8,112) (9,144) (10,180) (11,220)
        };
        \addlegendentry{Teledata D-C}
        
        % Telegate
        \addplot+[mark=square*, mark options={solid}, thick] plot coordinates {
            (2,2) (3,6) (4,12) (5,20) (6,30) (7,42) (8,56) (9,72) (10,90) (11,110)
        };
        \addlegendentry{Telegate D-C}
        
        % Expose
        \addplot+[mark=triangle*, mark options={solid}, thick] plot coordinates {
            (2, 2) (3, 5) (4, 9) (5, 14) (6, 20) (7, 27) (8, 35) (9, 44) (10, 54) (11, 65)
        };
        \addlegendentry{Expose D-C}

        \end{axis}
    \end{tikzpicture}

%% file: q_consumed_t-1c.tex
\begin{tikzpicture}
        \begin{axis}[
            xlabel={Number of QPUs connected},
            ylabel={Communication Qubits consumed},
            ymin=-5, ymax=260, 
            xmin=2, xmax=11,
            legend style={at={(0.25,0.95)}, anchor=north, draw=none, fill=none},
            legend cell align={left},
            xtick={1,2,3,4,5,6,7,8,9,10}
        ]
        
        % Teledata
        \addplot+[mark=*, mark options={solid}, thick] plot coordinates {
            (2, 8) (3, 20) (4, 36) (5, 56) (6, 80) (7, 108) (8, 140) (9, 176) (10, 216) (11, 260)
        };
        \addlegendentry{Teledata T-1C}
        
        % Telegate
        \addplot+[mark=square*, mark options={solid}, thick] plot coordinates {
            (2, 4) (3, 10) (4, 18) (5, 28) (6, 40) (7, 54) (8, 70) (9, 88) (10, 108) (11, 130)
        };
        \addlegendentry{Telegate T-1C}
        
        % Expose
        \addplot+[mark=triangle*, mark options={solid}, thick] plot coordinates {
            (2, 0) (3, 4) (4, 10) (5, 18) (6, 28) (7, 40) (8, 54) (9, 70) (10, 88) (11, 108)
        };
        \addlegendentry{Expose T-1C}

        \end{axis}
    \end{tikzpicture}

%% file: q_comm.tex
\begin{tikzpicture}
    \begin{axis}[
        xlabel={Number of QPUs connected},
        ylabel={Communication Qubits needed},
        ymin=0, ymax=30,
        xmin=2, xmax=11,
        legend style={at={(0.35,0.95)}, anchor=north, draw=none, fill=none},
        legend cell align={left},
        xtick={1,2,3,4,5,6,7,8,9,10}
    ]
    
    % Teledata
    \addplot+[mark=*, mark options={solid}, thick] plot coordinates {
        (2,4) (3,6) (4,8) (5,10) (6,12) (7,14) (8,16) (9,18) (10,20) (11,22)
    };
    \addlegendentry{Teledata D-C}
    
    % Telegate
    \addplot+[mark=square*, mark options={solid}, thick] plot coordinates {
        (2,2) (3,3) (4,4) (5,5) (6,6) (7,7) (8,8) (9,9) (10,10) (11,11)
    };
    \addlegendentry{Telegate D-C}
    
    % Expose
    \addplot+[mark=triangle*, mark options={solid}, thick] plot coordinates {
        (2, 1) (3, 1) (4, 1) (5, 1) (6, 1) (7, 1) (8, 1) (9, 1) (10, 1) (11, 1)
    };
    \addlegendentry{Expose, Telegate T-1C}

    % Teledata T1-Comm
    \addplot+[mark=diamond*, mark options={solid}, thick] plot coordinates {
        (2, 2) (3, 2) (4, 2) (5, 2) (6, 2) (7, 2) (8, 2) (9, 2) (10, 2) (11, 2)
    };
    \addlegendentry{Teledata T-1C}

    \end{axis}
\end{tikzpicture}

%% file: 5_Conclusions.tex
\section{Conclusions}\label{sec:conclusions}

This work addresses two key objectives aimed at mitigating some of the challenges identified in the literature regarding the development of compilation frameworks and software tools for distributed quantum computing. 

Firstly, the need to establish a common framework for the development of new IRs related to DQC is addressed by proposing the \emph{layered abstraction model}. This model not only provides a scalable architecture but also establishes a foundation for optimization opportunities in DQC. By implementing functions that facilitate quantum data distribution across logical topologies, NetQIR reduces the complexity of DQC programming, allowing compilers to dynamically optimise based on high-level semantic directives. This model effectively addresses challenges observed in other IRs, such as NetQASM and InQuIR, which either lack flexibility in protocol handling or are too closely tied to specific network assumptions.
  
Secondly, an \gls*{ir} has been proposed in this work that meets the requirements objectively specified in the abstraction layer model, called NetQIR. It is an innovative extension of QIR for \ac*{dqc}, designed to address current limitations in scalability and interoperability in distributed quantum environments. NetQIR offers a flexible \ac*{ir} designed to handle quantum and classical communications across multiple QPUs by introducing high-level abstractions and communication directives. Unlike previous solutions, NetQIR integrates high-level quantum communication functions ---such as point-to-point (\texttt{qsend}, \texttt{qrecv}) and collective operations (\texttt{scatter}, \texttt{gather}, \texttt{reduce}, \texttt{expose})--- enabling developers to program complex distributed algorithms with ease. This design abstracts the underlying network layer, allowing NetQIR to efficiently map communication protocols such as teledata and telegate based on the topology and specific requirements of the quantum network.

The quantitative results have allowed us to evaluate the quality of the abstraction layer by assessing that the semantic programming it provides yields better results while consuming fewer computational resources. Additionally, after demonstrating the quality of this model, we qualitatively evaluated the tools identified in related work. A quantitative evaluation of these tools remains challenging because many lack implementations. Moreover, it is important to consider that an \gls*{ir} is often just a simple specification by design.

Future work on NetQIR should prioritize developing tools that simplify its integration into new software projects. Additionally, testing its interoperability with various quantum backends and exploring advanced optimization techniques would be valuable. A well-designed toolchain could improve the management of distributed resources in NetQIR, potentially reducing communication costs and enhancing qubit allocation strategies to further boost efficiency and scalability in distributed quantum systems.